 \documentclass[12pt,epsf]{article}

\def\be{\begin{equation}}
\def\ee{\end{equation}}
\def\vp{\varphi}

\begin{document}
\begin{titlepage}
\bigskip
\rightline{}
\rightline{gr-qc/0507080}
\bigskip\bigskip\bigskip\bigskip
\centerline {\Large \bf {Higher Dimensional Generalizations of the Kerr Black Hole\footnote{To appear in {\sl Kerr Spacetime: Rotating Black Holes}, eds. S. Scott, M. Visser, and D. Wiltshire (Cambridge University Press).}}}
\bigskip\bigskip
\bigskip\bigskip

\centerline{\large  Gary T. Horowitz}
\bigskip\bigskip
\centerline{\em Department of Physics, UCSB, Santa Barbara, CA 93106}
\centerline{\em  gary@physics.ucsb.edu}
\bigskip\bigskip

\begin{abstract}

A brief introduction is given to rotating black holes in more than four spacetime dimensions.
\end{abstract}
\end{titlepage}

%\maketitle
%\chapter{Higher Dimensional Generalizations of the Kerr Black Hole}

\baselineskip 16pt
\setcounter{equation}{0}
\section{Introduction}
When  I was a graduate student at the University of Chicago in the late 1970's, I often heard Chandrasekhar raving about the Kerr solution \cite{Kerr:1963ud}. He was amazed by all of its remarkable properties and even its mere existence. As he said at the time: ``In my entire scientific life...the most shattering experience has been the realization that an exact solution of general relativity, discovered by the New Zealand mathematician Roy Kerr, provides the absolutely exact representation of untold numbers of massive black holes that populate the Universe" \cite{Chandra}.

It took me a while to understand Chandra's fascination, but I have come to agree. One can plausibly argue that the black hole solution discovered by Roy Kerr is the  most important vacuum solution ever found to Einstein's equation.
To honor Kerr's $70^{th}$ birthday, I would like to describe some recent generalizations of the Kerr solution to higher spatial dimensions. 

 Before I begin, let me say a word about the motivation for this work. There are two main reasons for studying these generalizations. The first comes from string theory, which  is a promising approach to quantum gravity. String theory predicts that spacetime has more than four dimensions. For a while it was thought that the extra spatial dimensions would be of order the Planck scale, making a geometric description unreliable, but it has recently been realized that there is a way to make the extra dimensions relatively large and still be unobservable. This is if we live on a three dimensional surface (a ``brane") in a higher dimensional space. String theory contains such higher dimensional extended objects, and it turns out that nongravitational forces are confined to the brane, but gravity is not. In such a  scenario,  all gravitational objects such as black holes are higher dimensional. The second reason for studying these solutions has nothing to do with string theory. Four dimensional black holes have a number of remarkable properties.  It is natural to ask whether these properties are general features of black holes or whether they crucially depend on the world being four dimensional. We will see that many of them are indeed special properties of  four dimensions and do not hold in  general.

\setcounter{equation}{0}
\section{Nonrotating  black holes in $D>4$}

To become familiar with black holes in higher dimensions, I will start by discussing  nonrotating black holes. (For a more extensive reviews of the material in this section, see  \cite{Kol:2004ww,Harmark:2005pp}.) For simplicity,  we will focus on $D=5$. There are two possible boundary conditions to consider: asymptotically flat in five dimensions, or the Kaluza-Klein choice -- asymptotically $M_4\times S^1$. In the asymptotically flat case, the only static black hole is the five dimensional Schwarzschild-Tangherlini solution \cite{Gibbons:2002av}
\be\label{5dsch}
ds^2 = - \left(1-{r_0^2\over r^2}\right) dt^2 +\left(1-{r_0^2\over r^2}\right)^{-1} dr^2
   + r^2 d\Omega_3 
\ee
  In the Kaluza-Klein case, there are more possibilities. Let $L$ be the length of the circle at infinity. The simplest solution with an event horizon is just the product of four dimensional Schwarzschild with radius $r_0$ and $S^1$: 
\be\label{blackstring}
ds^2 = - \left(1-{r_0\over r}\right) dt^2 +\left(1-{r_0\over r}\right)^{-1} dr^2
   + r^2 d\Omega + dz^2
\ee
This has horizon topology $S^2\times S^1$ and is sometimes called a black string, since it looks like a one dimensional extended object surrounded by an event horizon.  Gregory and Laflamme (GL) showed that this spacetime is unstable
to linearized perturbations with a long wavelength along the circle  \cite{Gregory:vy}. More
precisely, there is a critical size for the circle, $L_0$, of order $r_0$ such that
black strings with $L\le L_0$ are stable and those with $L>L_0$ are unstable.
The unstable mode is spherically symmetric, but
causes the horizon to oscillate in the $z$ direction.
Gregory and Laflamme also compared
the total entropy of the black string with that of a five dimensional spherical black hole
with the same total mass,
and found that when $L>L_0$, the black hole had greater entropy. They thus
suggested that the full nonlinear evolution of the instability would result
in the black string breaking up into separate black holes which would then
coalesce into a single black hole. Classically, horizons cannot bifurcate, but
the idea was that under classical evolution,
the event horizon would pinch off and become
singular. When the curvature became large enough, it was plausible that
quantum effects would smooth out the transition between the black string
and spherical black holes.

 However, it turns out that 
an event horizon cannot pinch off in finite time \cite{Horowitz:2001cz}. In particular, if one perturbs
(\ref{blackstring}), an $S^2$ on the horizon cannot  shrink to
zero size in finite affine parameter.
The reason is the following. Hawking's famous area theorem 
\cite{Hawking:1971tu}
is based on a local result that the divergence $\theta$ of the null geodesic
generators of the horizon cannot become negative, i.e., the null
geodesics cannot start to converge.
If an  $S^2$ on the horizon
tries to shrink to zero size, the null geodesics on that $S^2$ must be 
converging. The total $\theta$ can stay positive only if the horizon
is expanding rapidly in the circle direction, but this produces a large shear.
If the $S^2$ were to shrink to zero size in finite time, one can show
this shear would
drive $\theta$ negative. When it was realized that the black string cannot pinch off in finite time, it was suggested that the solution should settle down to a static nonuniform black string.

A natural place to start looking for these new solutions is with the static perturbation of the uniform black string that exists with wavelength $L_0$. Gubser \cite{Gubser:2001ac} did a perturbative calculation and found evidence that the nonuniform solutions with small inhomogeneity could not be the endpoint of the GL instability. 
Recent numerical work has found vacuum solutions describing static black strings with large inhomogeneity \cite{Wiseman:2002zc}. Surprisingly, all of these solutions have a mass which is larger than the unstable uniform black strings. So they cannot be the endpoint of the GL instability\footnote{In dimensions greater than 13 this changes and the nonuniform black string can be the endpoint of the instability \cite{Sorkin:2004qq}.}.  Solutions describing topologically spherical black holes in Kaluza-Klein theory have also been found numerically \cite{Kudoh:2003ki,Sorkin:2003ka}. When the black hole radius is much less than $L$, it looks just like (\ref{5dsch}).  As you increase the radius one finds that the size of the fifth dimension near the black hole grows and the black hole remains approximately spherical. It then reaches a maximum mass. Remarkably, one can continue past this point and find another branch of black hole solutions with lower mass and squashed horizons. It was conjectured by Kol \cite{Kol:2002xz} that the nonuniform black strings should meet the squashed black holes at a point corresponding to  a static solution with a singular horizon, and this appears to be the case \cite{Kudoh:2004hs}.  This yields a nice consistent picture of static Kaluza-Klein solutions with horizons, but it doesn't answer the question of what is the endpoint of the GL instability. An attempt to numerically evolve a perturbed black string is underway. An earlier attempt could not be followed far enough  to reach the final state \cite{Choptuik:2003qd}.

It was suggested by Wald \cite{Wald} that the black string horizon might pinch off in infinite affine parameter (avoiding the above no-go theorem), but still occur at finite advanced time as seen from the outside.  This is possible since the spacetime is singular when the horizon pinches off, and some evidence for this  has been found \cite{Marolf:2005vn}.
 If this were the case, then the original suggestion of Gregory and Laflamme that the black string will break up  into spherical black holes might still be correct.
 
\setcounter{equation}{0}
\section {Rotating black holes in $D>4$}

With Kaluza-Klein boundary conditions, the only known solution is the rotating black string obtained by taking the product of the Kerr metric and a circle.  Most of the recent work on higher dimensional rotating black holes has been in the context of asymptotically flat spacetimes, so from now on we will focus on this case.

The direct generalization of the Kerr metric to higher dimensions was found by Myers and Perry in 1986 \cite{Myers:1986un}.  In more than three spatial dimensions, black holes can rotate in different orthogonal planes, so the general solution has several angular momentum parameters. The general solution, with all possible angular momenta nonzero is known explicitly.  Like the Kerr metric, these solutions are all of the Kerr-Schild form \cite{KerrS} 
\be
g_{\mu\nu} = \eta_{\mu\nu} + h k_\mu k_\nu
\ee
where $k_\mu$ is null. 

If we set all but one of the angular momentum parameters to zero, we can write the metric in Boyer-Lindquist like coordinates. In $D$ spacetime dimensions, the solution is
$$
ds^2  = - dt^2 + \sin^2\theta (r^2 + a^2 ) d\vp^2 + {\mu\over r^{D-5} \rho^2} (dt - a\sin^2\theta d\vp)^2 $$
\be
+ \Psi^{-1} dr^2 + \rho^2 d\theta ^2 + r^2 \cos^2\theta d\Omega_{D-4}
\ee
where $\rho^2 = r^2 + a^2 \cos^2\theta$ and
\be
\Psi =  {r^2 + a^2\over \rho^2}  - {\mu\over r^{D-5}\rho^2}
\ee
Like the Kerr metric, this solution has two free parameters $\mu,a$ which determine the mass and angular momentum:
\be\label{MJ}
 M = {(D-2) \Omega_{D-2}\over 16\pi } \mu, \qquad J = {2\over D-2} Ma
 \ee
 where $\Omega_{D-2}$ is the area of a unit $S^{D-2}$.
One of the most surprising properties of these solutions is that for $D>5$, there is a regular horizon for all $M,J$! There is no extremal limit. This follows from the fact that the horizon exists where $\Psi=0$,  but for  $D>5$ this equation always has a solution. However, when the angular momentum is much bigger than the mass (using quantities of the same dimension, this is $J^{D-3} \gg M^{D-2}$) the horizon is like a flat pancake: it is spread out in the plane of rotation, but very thin in the orthogonal directions. Locally, it looks like  the product of $D-2$ dimensional Schwarzschild and $R^2$. It is probably subject to a GL instability \cite{Emparan:2003sy}. If so, there would be an effective extremal limit for stable black holes in all dimensions.

In five dimensions, there is an extremal limit, but the horizon area goes to zero in this limit. The extremal limit corresponds to $\mu=a^2$. Setting $D=5$ in (\ref{MJ}),  the mass and angular momentum are
\be 
M={3\pi\over 8}\mu\qquad\qquad J={2\over 3} Ma
\ee
so in the extremal limit
\be
J^2 = {32 \over 27 \pi} M^3
\ee

These solutions have recently been generalized to include a cosmological constant \cite{Hawking:1998kw,Gibbons:2004js}.

\setcounter{equation}{0}
\section{Black rings}

The above solutions all have a horizon which is topologically spherical $S^{D-2}$.
There is a qualitatively new type of rotating black hole that arises in $D=5$ (and possibly higher dimensions). One can take a black string, wrap it into a circle, and spin it along the circle direction just enough so that the gravitational attraction is balanced by the centrifugal force. Remarkably, an exact solution has been found describing this \cite{Emparan:2001wn}. It was not found by actually carrying out the above proceedure, but by a trick involving analytically continuing a Kaluza-Klein C-metric. The solution is
 independent of time, $t$, and two orthogonal 
rotations parameterized by $\varphi$ and $\psi$,  so the isometry group is $R\times U(1)^2$.  Introducing two 
other spatial coordinates, $-1\le x\le 1$ and $y\le -1$, the metric 
is:
$$ds^{2} = 
-\frac{F(y)}{F(x)}\left [dt + 
C(\nu)R \frac{1+y}{F(y)} d\psi\right ]^2
$$
\begin{equation}\label{blackring}
   + \frac{R^{2}}{(x-y)^{2}}F(x)  
    \Bigg [-\frac{G(y)}{F(y)} d\psi^{2} 
    -\frac{dy^2}{G(y)} +\frac{dx^2}{G(x)} 
    +\frac{G(x)}{F(x)} d\varphi^{2} \Bigg]
    \end{equation}
  where  
 $$
	    F(\xi) = 1 + {2\nu\over 1+\nu^2} \xi, \qquad  G(\xi) = 
	    (1-\xi^{2})(1+\nu \xi),
$$
 \be
 C(\nu) = {\nu\over 1+\nu^2}\left [{ 2(1+\nu)^3\over 1-\nu}\right]^{1/2}
 \ee
 and the angular variables are periodically identified with period
  \be
 \Delta\vp = \Delta \psi = {2\pi\over \sqrt{1+\nu^2}}
 \ee
 
 Although it is not obvious in these coordinates, this solution is asymptotically flat. The asymptotic region corresponds to $x\rightarrow -1, \ y\rightarrow -1$.
 The solution depends on two parameters $R>0$ and $0<\nu<1$ which determine the mass and angular momentum 
 \be
 M= {3\pi R^2\over 2}{\nu\over (1-\nu)(1+\nu^2)}, \qquad J = {\pi R^3\over 2} {C(\nu)\over(1-\nu)\sqrt{1+\nu^2}}
 \ee
  The coordinates $(x,\vp)$ parameterize two-spheres, so on a constant $t$ slice, surfaces of constant  $y<-1$ are topologically $S^2\times S^1$. The limiting value $y=-1$ corresponds to the axis for the $\psi$ rotations.
 The solution has an event horizon at $y=-1/\nu$  (where $G(y)=0$) with topology $S^2\times S^1$ and a curvature singularity at $y=-\infty$. It has no inner horizon, but it does have an ergoregion given by $-1/\nu < y < -(1+\nu^2)/2\nu$. This is just like the solution one would obtain by boosting the  black string (\ref{blackstring}) in the $z$ direction.
This solution is called a {\it black ring} rather than a black string since it is  wrapped around a  topologically trivial circle in space.

Not surprisingly, the angular momentum now has a lower bound, but no upper bound. Solutions exist only when $J^2 \ge M^3/\pi \equiv J_{min}^2$.  Something interesting happens when $J$ is near its  minimum value. If $J_{min}^2 < J^2 < {32\over 27} J_{min}^2$ there are two stationary black ring solutions with different horizon area. In addition, there is a Myers-Perry black hole! So for $M,J$ in this range, there are three black holes clearly showing the black hole uniqueness theorems do not extend to higher dimensions. Near $J_{min}$, the  Myers-Perry black hole has the largest horizon area. Near $(32/27)^{1/2} J_{min}$, one of the black rings has the largest area. 

It is not yet known if these solutions are stable, although the solutions with large $J$ are likely to be unstable.  This is because in this limit, the ring becomes  very long and thin, so one expects an analog of the GL instability. It cannot settle down to an inhomogeneous ring, since the rotation would cause the system to lose energy  due to  gravitational waves. The outcome of this instability (assuming it exists) is unknown.

\setcounter{equation}{0}
\section{Einstein-Maxwell solutions}

So far, we have considered only vacuum solutions.
The higher dimensional generalization of the Kerr-Newman solution is still not known analytically (but special cases have been found numerically \cite{Kunz:2005nm}). However exact solutions describing generalizations of the neutral black ring have been found. In fact, the
 discrete nonuniqueness discussed above becomes a continuously infinite nonuniqueness  when we add a Maxwell field. Suppose we consider  Einstein-Maxwell theory in five dimensions
\be
S=\int d^5x \sqrt{-g} \left (R- {1\over 4}F_{\mu\nu}F^{\mu\nu}\right )
\ee
In this theory, there is a global electric charge 
\be
Q ={1\over 4\pi} \int_{S^3} {}^*F
\ee
but no global magnetic charge since one cannot integrate the two-form $F$ over the three-sphere at infinity. However, there is a local magnetic charge carried by a string.
Given a point on a string, one can surround it by a two-sphere and define
\be
q= {1\over 4\pi} \int_{S^2} F
\ee
If the string is wrapped into a circle, there is no net magnetic charge at infinity. The asymptotic field resembles a magnetic dipole.  

It was suggested in  \cite{Reall:2002bh}  that there should exist a generalization of (\ref{blackring})
which has a third independent parameter labeling  the magnetic dipole moment $q$. The explicit solution was found about a year later \cite{Emparan:2004wy}. Note that unlike four dimensional  black holes, the dipole moment is an independent adjustible parameter. Since the only global charges are $M$ and $J$, there is continuous nonuniqueness. Solutions with $q\ne 0$ have a smooth inner horizon as well as an event horizon. There is now an upper bound on the angular momentum as well as a lower bound $J^2_{min} \le J^2 \le J^2_{max}$.  Solutions with $J^2= J^2_{max}$ are extremal in that the inner and outer horizons  coincide. The extremal solution has a smooth degenerate horizon, with zero surface gravity. 

This dipole charge enters the first law in much the same way as an ordinary global charge. There is a corresponding potential $\phi$ and one can show \cite{Copsey:2005se} that for any perturbation satisfying the linearized constraints
 \be
\delta M = \frac{\kappa}{8\pi} \delta{A}_{H} + \Omega_{i} 
\delta{J}^{i} + \Phi \delta Q+ \phi 
\delta q 
 \ee
 where, as usual,  $\kappa$ is the surface gravity, $A_H$ is the horizon area, $\Omega_i$ ($i=1,2$) are the angular velocities in the two orthogonal planes, and $\Phi$ is the electrostatic potential.
 We have included the possibility of a global electric charge  and rotation in both orthogonal planes even though the dipole ring found in  \cite{Emparan:2004wy} has zero electric charge and only one nonzero angular momentum.

\setcounter{equation}{0} 
\section{Supersymmetry}

Let us first recall the situation with supersymmetric black holes in four dimensions. In $D=4$, supersymmetry requires $M=Q$, so the only supersymmetric black holes in Einstein-Maxwell theory are
extremal Reissner-Nordstr\"om and superpositions of them. The Kerr-Newman solution with $M=Q$, and $J\ne 0$ is supersymmetric but describes a naked singularity. (If one adds additional matter fields, there exist supersymmetric multi black hole solutions with angular momentum \cite{Bates:2003vx}.)

The situation is different in $D=5$. To begin, the bosonic sector of the minimal supergravity theory is not just Einstein-Maxwell, but also contains a Chern-Simons term
\begin{equation}\label{sugra}
    S = \int d^5 x\sqrt{-g} \Big(R -\frac{1}{4} F_{\mu \nu} 
F^{\mu \nu} -{1\over 12\sqrt 3} 
    \epsilon^{\mu \nu \rho \sigma \eta} F_{\mu \nu} F_{\rho \sigma} 
A_{\eta} \Big)
    \end{equation}
    Surprisingly, the addition of the Chern-Simons term makes it easier to solve the field equations\footnote{This is because one can employ various solution generating techniques known for supergravity.},  and analytic solutions describing charged,  rotating black holes in this theory are known \cite{Breckenridge:1996sn}.
Their extremal limit provides a two parameter family of rotating supersymmetric black holes \cite{Breckenridge:1996is}. One parameter determines the angular momenta: the angular momenta in both orthogonal planes are nonzero and equal, $J_1=J_2$. The second determines the electric charge.  The mass is determined in terms of the charge $M=(\sqrt 3/2)Q$.  As expected, the surface gravity vanishes  for these extremal solutions, but surprisingly, the angular velocity also vanishes. The fact that there is no ergoregion follows from the fact that supersymmetric, asymptotically flat solutions must have a Killing field that remains timelike (or null) everywhere outside the horizon \cite{Gauntlett:1998fz}.  
These horizons are topologically $S^3$, and Reall has shown that these are the only supersymmetric solutions with spherical horizons \cite{Reall:2002bh}.

However, there are also supersymmetric black rings with $S^2\times S^1$ horizons \cite{Elvang:2004rt}.  The dipole rings discussed above are also solutions to (\ref{sugra}) since the Chern-Simons term does not contribute in this case, but they are not supersymmetric. To obtain supersymmetric solutions, one must add electric charge and take a limit so that $M=(\sqrt 3/2)Q$.  It turns out that the supersymmetric solutions can have  two independent angular momenta, so there are three  free parameters $Q,J_1,J_2$. There is also a  magnetic dipole  charge but it is  not independent. 

The near horizon limit of a $D=4$ extreme Reissner-Nordstrom black hole is $AdS_2\times S^2$ (where AdS denotes anti de Sitter spacetime). Similarly, the near horizon limit of the supersymmetric black ring is $AdS_3\times S^2$. Interestingly enough, the near horizon limit of the  extreme Kerr metric  and some Myers-Perry solutions also resemble (warped) products of  AdS  and a sphere \cite{Bardeen:1999px}.

One of the advantages of considering supersymmetric solutions is that they minimize the mass for given charges.  Since there is no lower mass solution to decay to, these supersymmetric solutions are expected to be stable. Another advantage is that it is easier to count their microstates in string theory and compare with the 
Hawking-Bekenstein entropy. This was done successfully for the topologically spherical black hole \cite{Breckenridge:1996is} and has also been discussed for the black rings \cite{Cyrier:2004hj,Bena:2004tk}.
  
 In this example of five dimensional minimal supergravity, the supersymmetric solution is uniquely determined by the conserved charges at infinity. However, in ten dimensional  supergravity, supersymmetric solutions have been found with three types of electric charges and three local magnetic charges but only one constraint \cite{EEMR,Gauntlett:2004wh,Bena:2004de}. So together with the two angular momenta there are a total of seven parameters, but only five global charges. So even among supersymmetric solutions, there is continuous nonuniqueness.

\setcounter{equation}{0}
\section{Discussion}
 In four dimensions, black holes enjoy a number of special properties including:

 1) They are uniquely specified by $M,J,Q$
 
 2) The are topologically  spherical
 
 3) They are stable
 
\noindent We have seen that none of these properties are preserved in higher dimensions. The solutions can be labelled by local charges rather than global charges.  In five dimensions, black hole horizons can have topology $S^2\times S^1$ as well as $S^3$. Finally, we have seen that the straight black string is unstable, and highly rotating black rings and black holes in more than five dimensions are likely to be also. With regard to the nonuniqueness, it should be noted that there is still a finite number of parameters characterizing the solution. When this is compared with the vast number of ways of forming these objects, one can take the view that the spirit  of the no hair theorem is preserved\footnote{I thank B. Carter for stressing this point.}.

A surprisingly large number of higher dimensional generalizations of the Kerr solution have been found. But the list is far from complete, and a great deal remains to be understood.  For example, we do not yet have good restrictions on the topology of stationary black holes in $D>4$. Clearly $S^n$ and $S^2\times S^1$ are possible. Are there others? There should also be more general black ring solutions in five dimensions. The vacuum solution given in (\ref{blackring}) has angular momentum only along the circle. One expects that it should also be possible to have angular momentum on the two-spheres. This would be analogous to first taking the product of the $D=4$ Kerr metric and a line to obtain a rotating black string, and then wrapping the string into a circle and spinning it to obtain another stationary black ring. Finally, all of the known higher dimensional black holes have a great deal of symmetry: in addition to time translation invariance, they are all invariant under rotations in mutually orthogonal planes. It was pointed out in \cite{Reall:2002bh} that on general grounds, one only expects rotating black holes to have one additional Killing field. So there may well exist much more general black holes in $D>4$ with only two Killing fields. 

In short,  the space of black holes is much richer in more than four spacetime dimensions. It remains to be seen whether nature takes advantage of this richness.

\vskip 1cm
\centerline{\bf Acknowledgements}
\vskip .5cm
I would like to thank the organizers of the Kerr Fest (Christchurch, New Zealand, Aug. 26-28, 2004) for a stimulating meeting. I also want to thank H. Elvang and H. Reall for their explanations of some the results discussed here. This work was supported in part by NSF grant  PHY-0244764.

\end{document}